\documentclass[hyper]{JHEP} 

\usepackage{epsfig}
\usepackage{amsfonts}
\usepackage{graphicx}

\def\be{\begin{equation}}
\def\ee{\end{equation}}
\def\ba{\begin{array}}
\def\ea{\end{array}}
\def\bea{\begin{eqnarray}}
\def\eea{\end{eqnarray}}

\def\ba{\mathbf{a}}

\newcommand\fverb{\setbox\pippobox=\hbox\bgroup\verb}

\newcommand\fverbdo{\egroup\medskip\noindent%

            \fbox{\unhbox\pippobox}\ }

\newcommand\fverbit{\egroup\item[\fbox{\unhbox\pippobox}]}

\newbox\pippobox

\title{Fundamental String (Membrane) Orbiting 
D5(M5)-branes}
\author{Rashmi R. Nayak\\
Centre for Theoretical Studies \\
Indian Institute of Technology Kharagpur, \\
Kharagpur- 721 302, India\\
E-mail: \email{rashmi@cts.iitkgp.ernet.in}}
\author{Pratap K. Swain\\
Department of Physics \& Meteorology \\
Indian Institute of Technology Kharagpur, \\
Kharagpur, 721 302, India \\
E-mail: \email{pratap@phy.iitkgp.ernet.in}}

\preprint{{yymm.nnnn[hep-th]}}

\abstract
{We study fundamental string (F-string) dynamics near $D5$-brane
in some limits. 
We find that when the angular momentum of the probe is 
proportional to string length $(l_s)$ and square root of string 
coupling constant $(\sqrt{g_s})$, the F-string lies in its 
metastable orbit at a finite distance from $D5$-branes. 
We further study the metastable orbits of a M2-brane in 
M5-brane background.}
\def \ba{\mathbf{a}}

\begin{document}

\section{Introduction and Summary}\label{first}
Type II string theories admit, in addition to the 
usual BPS branes,
non-BPS branes \cite{Sen:1999mg} in its spectrum which are unstable due to the presence of 
the open string tachon. By now 
it has been well understood that the
dynamics of this open string tachyon can be 
described by a Dirac-Born-Infeld
(DBI) type of action which captures all the essential properties of
the tachyon field \cite{Sen:2002nu}, \cite{Sen:2002in},  \cite{Sen:2002an}, \cite{Sen:2002vv}, \cite{Sen:2003tm}, \cite{Kutasov:2003er}. For a comprehensive
review and complete list of references refer to \cite{Sen:2004nf}. The prefactor that appears in front of the Lagrangian
density is the tachyon potential. Not so long back, 
it was noticed that
the open string tachyon dynamics on the non-BPS 
brane has a geometric meaning
in terms of the rolling of a BPS D-brane 
in the vicinity of a stack 
of NS5-branes, the so-called geometric tachyon \cite{Kutasov:2004dj}, \cite{Kutasov:2004ct}. 
This is nothing but the
radial distance between the probe D-brane 
and the NS5-brane, and the
time dependent dynamics of this D-brane 
has shown to be astonishingly 
similar to that of the open string tachyon dynamics. 
As the D-brane moves in the vicinity of the 
gravitational potential produced by the NS5-brane
\footnote{the Harmonic function for a stack of NS5-branes goes like 
$\frac{1}{r^2}$}, it pulls it towards its core 
and finally it gets 
absorbed into the NS5-brane \cite{Kutasov:2004dj}. For the 
energy of the incoming D-brane $E>\tau_p$, the D-brane escapes to infinity and for $E<\tau_p$, 
the D-brane will fall into the NS5-brane. \footnote{Further studies on
rolling branes into NS5-brane have been considered, for example, in \cite{Nakayama:2004yx}, \cite{Sahakyan:2004cq},
 \cite{Kluson:2004xc}, \cite{Chen:2004vw}, \cite{Nakayama:2004ge}.}

However, more recently, in \cite{Jun:2009bn} it was
argued that for a particular value of energy and angular momentum of the probe D-brane, the brane will
lie in a metastable orbit and revolve around the NS5-brane by keeping a certain distance from NS5-brane at all times. In this limit of energy and angular momentum, the induced tachyon field on D-brane becomes a massless field due to 
the constant tachyon potential and the D-brane becomes a stable object. In this paper \cite{Jun:2009bn}, this mechanism was well explained by constructing an classical equivalent radial 
action involving angular momentum. It was argued that the NS5-brane has no effecft on orbiting D-brane but the observer on the D-brane sees the change 
in tension due to NS5-brane.

In an attempt to understand the origin of  this geometric 
tachyon in \cite{Das:2008af} the time dependent 
dynamics of probe D-brane in the background of NS5-brane 
was explained in terms of proper acceleration. It was 
proposed that the tachyonic instability is due to the geodesic deviation caused by 
proper acceleration (which is formed due to the background dilaton field). This idea has been further extended to 
the system of F-string falling into the $Dp$-brane  \cite{Das:2008gi}, even though there was no dilaton prefactor in the Nambo-Goto action for F-string in $Dp$-brane background. It was found that the tachyonic instability in (F-$D5$) system is due 
to the over all conformal factor of the induced metric on  the F-string. Perhaps, this was in the expected line 
as (F-$D5$) system is S-dual to the ($D1$-NS5)  and the later has a tachyonic instability due to the geometric tachyon. The resemblance of the (F-$D5$) system with 
that of open string dynamics on the non-BPS brane, it is 
interesting to study the F-string orbiting around 
the stack of $Dp$-branes. We shall study this system 
in the present paper.

Furthermore, in a related paper \cite{Xu:2007wj}
 the dynamics of the M2-brane was studied near the 
M5-brane background. As it is well known that M2-brane and 
M5-brane in 11-dimensions can be reduced to 
D-branes or other branes in string theory by 
compactifying some directions and applying T-dualities. 
Both M2-branes and M5-branes are stable and preserving some 
supersymmetries, but M2-brane in the vicinity of M5-brane 
breaks all supersymmetries. It was shown in \cite{Xu:2007wj}
 that 
the dynamics of M2-brane in the background of M5-brane behaves similarly as that of D-brane in the 
vicinity of NS5-brane after compactifying one of the 
transverse direction ($x^{11}$) at a periodic interval of 
$2{\pi}R_{11}$ with limit $1<<\frac{r}{R}$, where $r$ is 
the radius of M5-brane along the transverse directions.

Motivated by the recent study of D-brane orbiting in NS5-brane background, we study further examples of such motion in various string and M-theory backgrounds. As specified earlier, the (F-$Dp$) system behaves in the 
same way as the open string tachyon condensation on 
non-BPS brane, we study the dynamics of F-string in the 
orbiting limit around a stack of $D5$-branes background. 
We find that when the angular momentum of the incoming 
F-string is related to the string coupling $(g_s)$ and 
the string length $(l_s)$, the F-string will lie in its 
metastable orbit keeping certain distance from the 
$D5$-brane. We further propose an action for the F-string 
written in terms of the radial coordinate only which 
gives the same equations of motion and conserved 
quantities but with a different tachyon-like potential 
compared to the one proposed in \cite{Das:2008gi}. Then 
we study the dynamics of a $Dp^{\prime}$-brane in the 
background of a stack of $Dp$-branes in the orbiting 
limit. \footnote{We have chosen the probe and the 
background in such a way that there is no Wess-Zumino 
term in the action of the $Dp$-brane and the dynamics is governed by the 
dilaton and by the metric.} 
It so happens that only the $(D3-D5)$ system 
shows a similar behavior in orbiting limit with 
non-zero angular momentum.

Finally, we study the M2-brane dynamics in a stack of 
M5-branes background in orbiting limit. We show that 
there exists a solution where the angular momentum 
is written in terms of Planck length and the size of the 
eleven dimensional circle where the M2-brane will lie in 
the metastable orbit and revolve around the M5-brane 
by keeping a certain distance  from it. Further we 
write the action of such a M2-brane, in terms of 
radial coordinate only, which gives the same equations 
of motion and conserved quantities, but with an 
effective tension.

The rest of the paper is organized as follows. In section 
\ref{second}, we study the F-string orbiting around the $D5$-brane 
background. Section \ref{third} is devoted to the study of 
orbiting limit of $Dp^{\prime}$-brane in $Dp$-brane 
background. Finally, in section \ref{fourth}, we study the M2-brane 
in M5-brane background in orbiting limit.

\section{F-string Orbiting a stack of D5-branes}\label{second}
(F-$D5$) is a non-supersymmetric system because they break 
different halves of supersymmetries. Therefore, F-string 
in the vicinity of $D5$-branes form a non-BPS system. The dynamics of F-string in $D5$-brane background has been 
discussed in \cite{Das:2008gi}. We wish to find the condition where F-string is orbiting around $D5$-brane. 
The supergravity solutions of parallel coincident $D5$-branes are given by:

\begin{eqnarray}
ds^2 &=& {H(r)}^{-\frac{1}{2}}\bigg(-dt^2 + 
\sum_{i=1}^5(dx^i)^2\bigg)
+ {H(r)}^{\frac{1}{2}}\bigg(dr^2 + d{\Omega_3}^2\bigg),
\cr & \cr
e^{2(\phi - \phi_0)} &=& {H(r)}^{-1}, \cr & \cr
H(r) &=& 1 + \frac{Ng_s{l_s}^2}{r^2}, \cr & \cr
F_7 &=& d{H(r)}^{-1}\wedge dt\wedge dx^1\wedge...\wedge dx^5.
\nonumber \\ \label{12}
\end{eqnarray}
Here $H(r)$ is the harmonic function and $r=\sum_{a=6}^9(dx^a)^2$, 
$\phi$ is the dilaton, $d{\Omega_3}$ is the volume element 
of 3-sphere in transverse directions of $D5$-branes and 
$F_7$ is the 7-form Ramond-Ramond (R-R) field strength. The Nambo-Goto action of F-string is given by:
\begin{equation}
S_f = -\frac{1}{2\pi {l_s}^2}\int d^2\xi\sqrt{-{\rm det}~{\rm G}_{\mu\nu}}
\nonumber \\ \label{13}
\end{equation}
where, $\mu,\nu$ runs from 0 to 1. 
and $d^2\xi$ represents the area element of 
the world sheet of F-string.  $G_{\mu\nu}$ is the induced 
metric on the string given by
\begin{equation}
G_{\mu\nu} = \frac{\partial X^A}{\partial{\xi}^{\mu}}
\frac{\partial X^B}{\partial{\xi}^{\nu}}G_{AB},
\nonumber \\ \label{14}
\end{equation}
where $A,B = 0,1,...9$. We set, by reparametrization, $\xi^{\mu} = X^{\mu}$.
The position of the F-string in the transverse space of the $D5$-brane
gives rise to scalars on the world volume of the string. We further 
presume that the transverse directions are function
of time $(t)$ only. Under this, the
action (\ref{13}) for the string lying in $(t, x^1)$ is given by:
\begin{equation}
S_f = -\frac{1}{2\pi {l_s}^2}\int dt dx^1 \frac{1}
{\sqrt H(X^m)}
\sqrt{1 - H(X^m){\dot{X}}^m{\dot{X}}^m}.
\nonumber \\ \label{15}
\end{equation}
The Lagrangian is given by
\begin{equation}
{\cal L} =  -\frac{1}{2\pi {l_s}^2}\frac{1}{\sqrt H(X^m)}
\sqrt{1 - H(X^m){\dot{X}}^m{\dot{X}}^m},
\nonumber \\ \label{16}
\end{equation} 
and the nonzero components of stress-energy tensor   
$T_{\mu\nu}$ are given by:
\begin{equation}
T_{00} = \tau_f\frac{{H(X^m)}^{-\frac{1}{2}}}
{\sqrt{1 - H(X^m){\dot{X}}^m{\dot{X}}^m}},
\nonumber \\ \label{17}
\end{equation}
\begin{equation}
T_{ij} = -\tau_f{H(X^m)}^{-\frac{1}{2}}
\sqrt{1 - H(X^m){\dot{X}}^m{\dot{X}}^m}\delta_{ij},
\nonumber \\ \label{18}
\end{equation}
where $\tau_f = \frac{1}{2\pi {l_s}^2}$ is the fundamental 
string tension. The energy density $(E)$ 
and angular momentum  $(L)$ are defined as
\begin{eqnarray}
E &=& P_n{\dot{X}}^n - \cal L  \cr & \cr
L &=& X^mP^n - X^nP^m,
\nonumber \\ \label{19}
\end{eqnarray}
where, $P_n$ is the conjugate momentum defined by:
\begin{equation}
P_n = \frac{\delta\cal L}{\delta{\dot{X}}^n} =
\tau_f\frac{{H(X^m)}^{\frac{1}{2}}
{\dot{X}}^n }{\sqrt{1 - H(X^m){\dot{X}}^m{\dot{X}}^m}}.
\nonumber \\ \label{20}
\end{equation}
Using $P_n$ as given in (\ref{20}) in equation (\ref{19}), 
the angular momentum and energy density becomes
\begin{equation}
L = \tau_f\frac{{H(X^m)}^{\frac{1}{2}}}
{\sqrt{1 - H(X^m){\dot{X}}^m{\dot{X}}^m}}
\bigg[X^m{\dot{X}}^n - X^n{\dot{X}}^m\bigg],
\nonumber \\ \label{21}
\end{equation}
\begin{equation}
E = \tau_f\frac{{H(X^m)}^{-\frac{1}{2}}}
{\sqrt{1 - H(X^m){\dot{X}}^m{\dot{X}}^m}}.
\nonumber \\ \label{22}
\end{equation}
If F-string is always confined in the $(X^6,X^7)$ plane, using polar 
coordinates as $X^6 = R\cos\theta$
and $X^7 = R\sin\theta$, the energy density and  the angular momentum becomes 
\begin{equation}
E = \tau_f\frac{{H(R)}^{-\frac{1}{2}}}
{\sqrt{1 - H(R)({\dot{R}}^2 +
R^2{\dot{\theta}}^2)}},
\nonumber \\ \label{23}
\end{equation}
and
\begin{equation}
L = \tau_f\frac{{H(R)}^{\frac{1}{2}}R^2\dot{\theta}}
{\sqrt{1 - H(R)({\dot{R}}^2 +
R^2{\dot{\theta}}^2)}}.
\nonumber \\ \label{24}
\end{equation}
The equations for ${\dot{\theta}}$ and ${\dot{R}}$ 
derived from the above two equations become
\begin{eqnarray}
{\dot{\theta}}^2 &=& \frac{l^2}{{\epsilon}^2R^4{H(R)}^2},
\cr & \cr {\dot{R}}^2 &=& \frac{1}{{\epsilon}^2{H(R)}^2}
\Bigg[{{\epsilon}^2H(R) - \bigg(1 + \frac{l^2}{R^2}\bigg)
}\Bigg].
\nonumber \\ \label{25}
\end{eqnarray}
Where we have defined $\epsilon = \frac{E}{\tau_f}$ and
$l = \frac{L}{\tau_f}$.
This radial equation can be compared with a particle 
of mass $(m=2)$ moving in a one dimensional effective potential (with zero energy) as:
\begin{equation}
V_{eff} = -\frac{1}{{\epsilon}^2{H(R)}^2}
\Bigg[{{\epsilon}^2H(R) - \bigg(1 + \frac{l^2}{R^2}
\bigg)}\Bigg].
\nonumber \\ \label{26}
\end{equation}
Note that the effective potential vanishes for 
$\epsilon =1$ and $l = \sqrt{Ng_s}l_s$. Similar type of calculation has been 
done for the D-brane orbiting around the NS5-branes in \cite{Jun:2009bn}. The orbiting angular momentum 
depends both on 
string length $(l_s)$ and string coupling constant $(g_s)$. In analogy with \cite{Jun:2009bn}, we conclude 
that F-string maintains
a particular desired orbit around the $D5$-brane
as long as the above energy and angular momentum limits are satisfied, because
for vanishing $V_{eff}$, there is no force that pulls the string towards the 5-brane 
or pushes it to infinity.  

As in \cite{Jun:2009bn}, we can write the action given in (\ref{15}) in terms of polar coordinates as
\begin{equation}
S_f = -\tau_f\int dtdx^1{{H(R)}^{-\frac{1}{2}}}
{\sqrt{1 - H(R)({\dot{R}}^2 +
R^2{\dot{\theta}}^2)}}.
\nonumber \\ \label{32}
\end{equation}
The equations of motion for $R$ and $\theta$ derived from 
this action are
\begin{equation}
\frac{d}{dt}\Bigg(\frac{\dot{R}{H(R)}^{\frac{1}{2}}}
{\sqrt{1-H(R)\big({\dot{R}}^2 + 
R^2{\dot{\theta}}^2\big)}}\Bigg) = 
\frac{R{H(R)}^{\frac{1}{2}}}
{\sqrt{1-H(R)\big({\dot{R}}^2 + 
R^2{\dot{\theta}}^2\big)}}\Bigg[{\dot{\theta}}^2 -
{\omega^2(R)}\Bigg],
\nonumber \\ \label{33}
\end{equation}
and
\begin{equation}
\frac{d}{dt}\Bigg(\frac{R^2\dot{\theta}
{H(R)}^{\frac{1}{2}}}
{\sqrt{1-H(R)\big({\dot{R}}^2 + 
R^2{\dot{\theta}}^2\big)}}\Bigg) = 0.
\nonumber \\ \label{35}
\end{equation}
Where $\omega(R) = \frac{\sqrt{N g_s} l_s}{R^2 H(R)}$. 
One can check that $\dot{R} = 0$ with $\epsilon =1$
and $l=\sqrt{Ng_s}l_s$ becomes a solution to
(\ref{33}). Further the equation (\ref{35}) guaranty 
that $L$ is a constant of motion.

Now, we consider an equivalent radial action 
involving angular momentum as 
\begin{equation}
{S}^{\prime}_f = -\tau_f\int dt\sqrt{1+\frac{l^2}{R^2}}
\sqrt{\frac{1}{H(R)}- {\dot{R}}^2}.
\nonumber \\ \label{39}
\end{equation}
Though this action looks same as given in  \cite{Jun:2009bn}, 
the expressions for $H$ and $l$ are not same. From the 
equivalent action, we get same radial equation of motion 
and energy density as that of original action (\ref{32}). By 
comparing the action (\ref{39}) with the open 
string tachyon effective action given in \cite{Sen:2002nu}, we get the tachyon 
potential as $V(T) = \tau_f\frac{\sqrt{1+
\frac{l^2}{R^2}}}{H(R(T))}$ which is different than that 
given in \cite{Das:2008gi}.  Note that for the orbiting condition, $l = \sqrt{N g_s} l_s$, the tachyon potential $\tilde{V} (T) = \tau_f$.
The geometric tachyon does not roll, because the tachyon
potential becomes flat. The geometric tachyon induced on the string becomes a massless scalar as described in \cite{Das:2008gi}. Hence the decay process is suppressed 
and the F-string becomes stable.  

\section{$Dp^{\prime}$-brane Orbiting  
$Dp$-branes}\label{third}

The supergravity solution of a stack of $N$ 
coincident $Dp$-branes is given by the following form of 
metric, dilaton $(\phi)$, and R-R field $C_{(p+1)}$ as
\begin{eqnarray}
ds^2 &=& {H_p}^{-\frac{1}{2}} 
\sum_{a=0}^p(dx^a)^2 + 
+ {H_p}^{\frac{1}{2}}
\sum_{i=p+1}^9(dx^i)^2,\cr & \cr
e^{2\phi} &=& {H_p}^{\frac{3-p}{2}}, \cr & \cr
H_p &=& 1 + \frac{Ng_s{l_s}^{7-p}}{r^{7-p}}, \cr & \cr
C_{0 \cdots p} &=& {H_p}^{-1}.
\nonumber \\ \label{99}
\end{eqnarray}
Where $H_p$ is the harmonic function of $N$ coincident 
$Dp$-branes in the transverse directions of $Dp$-brane. In the subsequent analysis, we shall assume that (1) the dimensionality of the
probe is smaller than the background, (2) there is
no magnetic flux on the probe, (3) we shall take a single probe brane at 
a time. Then the dynamics will be given by the induced metric and the
dilaton prefactor only. 
The action of such a probe $D{{p}^{\prime}}$ in a $Dp$
brane background is given by DBI action as in \cite{Panigrahi:2004qr}
\begin{equation}
S_{{p}^{\prime}} = -\tau_{{p}^{\prime}}
V\int dt{H_p}^{\frac{p -
{p}^{\prime} -4}{4}}\sqrt{1 - {\dot{X}}^m{\dot{X}}^mH_p}.
\nonumber \\ \label{98}
\end{equation}
To get the action (\ref{98}), we set the 
reparametrization invariance of the world volume 
coordinates of $Dp^{\prime}$-brane 
(leveled as ${\xi}^a = x^a$), gives rise scalar 
fields $(X^m)$ along the transverse directions 
of D-brane. We restrict the radial fluctuations 
along the transverse directions 
$(R = \sqrt{X^mX_m({\xi}^a)})$ only and also the scalar 
fields are the function of time $(t)$.

The nonzero components of stress-energy tensors $T_{\mu\nu}$ are given by
\begin{equation}
T_{00} = \tau_{{p}^{\prime}}\frac{{H_p}^{\frac{p -{p}^
{\prime} -4}{4}} }{\sqrt{1 - {\dot{X}}^m{\dot{X}}^mH_p}},
\nonumber \\ \label{1}
\end{equation}
\begin{equation}
T_{ij} = -\tau_{{p}^{\prime}}{H_p}^{\frac{p -{p}^
{\prime} -4}{4}}\sqrt{1 - {\dot{X}}^m
{\dot{X}}^mH_p}\delta_{ij},
\nonumber \\ \label{2}
\end{equation}
where we have taken $\int{d^p x} = V = 1$. 
The conserved quantities are energy $(E)$ and
angular momentum $(L)$ and they are defined as
\begin{eqnarray}
E &=& P_n{\dot{X}}^n - \cal L, \cr & \cr
L &=& X^mP^n - X^nP^m.
\nonumber \\ \label{3}
\end{eqnarray}
Where the conjugate momentum $P_n$ is given by:
\begin{equation}
P_n = \frac{\delta\cal L}{\delta{\dot{X}}^n} =
\tau_{{p}^{\prime}}\frac{{H_p}^{\frac{p -{p}^
{\prime}}{4}}{\dot{X}}^n }{\sqrt{1 -
{\dot{X}}^m{\dot{X}}^mH_p}}.
\nonumber \\ \label{4}
\end{equation}
The angular momentum and the energy now becomes 
\begin{equation}
L = \tau_{{p}^{\prime}}\frac{{H_p}^{\frac{p -{p}^
{\prime}}{4}}}{\sqrt{1 - {\dot{X}}^m{\dot{X}}^mH_p}}
\bigg[X^m{\dot{X}}^n - X^n{\dot{X}}^m\bigg],
\nonumber \\ \label{5}
\end{equation}
and
\begin{equation}
E = \tau_{{p}^{\prime}}\frac{{H_p}^{\frac{p -{p}^
{\prime} - 4}{4}}}{\sqrt{1 - {\dot{X}}^m{\dot{X}}^mH_p}}.
\nonumber \\ \label{6}
\end{equation}
Considering the motion of the $Dp^{\prime}$ lies in the 
plane $(X^6,X^7)$ at all times and using polar 
coordinates $X^6 = R\cos\theta$
and $X^7 = R\sin\theta$, the angular
momentum and energy becomes
\begin{equation}
E = \tau_{{p}^{\prime}}\frac{{H_p}^{\frac{p -{p}^
{\prime} - 4}{4}}}{\sqrt{1 - ({\dot{R}}^2 +
R^2{\dot{\theta}}^2)H_p}},
\nonumber \\ \label{7}
\end{equation}
and
\begin{equation}
L = \tau_{{p}^{\prime}}\frac{{H_p}^{\frac{p -{p}^
{\prime}}{4}}R^2\dot{\theta}}
{\sqrt{1 - ({\dot{R}}^2 +
R^2{\dot{\theta}}^2)H_p}}.
\nonumber \\ \label{8}
\end{equation}
Solving the above two equtions, for $\dot{R}$ and
$\dot{\theta}$, we get
\begin{eqnarray}
{\dot{\theta}}^2 &=& \frac{l^2}{{\epsilon}^2R^4{H_p}^2},
\cr & \cr {\dot{R}}^2 &=& \frac{1}{{\epsilon}^2{H_p}^2}
\Bigg[{{\epsilon}^2H_p - \bigg({H_p}^
{\frac{p-p^{\prime}-2}{2}}
+ \frac{l^2}{R^2}\bigg)}\Bigg].
\nonumber \\ \label{9}
\end{eqnarray}
Where once again we have defined $\epsilon = \frac{E}{\tau_{{p}^{\prime}}}$ and
$l = \frac{L}{\tau_{{p}^{\prime}}}$.
The radial equation of motion describes a particle 
of mass $(m=2)$ moving
in a one dimensional effective potential
(with zero energy) as
\begin{equation}
V_{eff} = -\frac{1}{{\epsilon}^2{H_p}^2}
\bigg[{{\epsilon}^2H_p - \bigg({H_p}^
{\frac{p-p^{\prime}-2}{2}}
+ \frac{l^2}{R^2}\bigg)}\bigg].
\nonumber \\ \label{10}
\end{equation}
Note that the effective potential vanishes for
\begin{eqnarray}
\epsilon &=& 1, \cr & \cr
p &=& p^{\prime} + 2, \cr & \cr {\rm and} ~~~ 
H_p &=& 1 + \frac{l^2}{R^2}.
\nonumber \\ \label{11}
\end{eqnarray}
The last two conditions together tell that $V_{eff}$ vanishes only when
$p = 5, p^{\prime} = 3$ and $l = \sqrt{Ng_s}l_s$. Thus the only consistent orbiting 
condition is found in case of $D3$-brane in $D5$-brane background. Rest of the analysis can be done by replacing $l = \sqrt{Ng_s}l_s$ and $H = 
1+\frac{N g_s l^2_s}{R^2}$ in \cite{Jun:2009bn} and 
in section \ref{second} of this paper. It will be interesting to find out further example of $Dp$-brane orbiting
around stacks of D-brane bound states. 

\section{ Membrane Orbiting M5-brane}
\label{fourth}
In this section, we shall study the membranes orbit in a background generated
by a periodic configuration of $N$ coincident M5-branes along the $x^{11}$
direction at an intervals $2\pi R_{11}$. In the limit of $1
<< r/R_{11}$, the background of such array of M5-branes becomes \cite{Xu:2007wj}, \cite{Gueven:1992hh} 
\bea
 ds^2&=& H^{-\frac{1}{3}} \eta_{\mu\nu} dx^\mu dx^\nu
 + H^{\frac{2}{3}}\delta_{ij} dx^i dx^j+
  H^{\frac{2}{3}}(dx^{11})^2,
  \nonumber \\
    H &=& 1 + \frac{N l_p^3}{R_{11}r^2},  \nonumber \\
    F_4 &=& \frac{2N\ell_p^3}{R_{11}} dv_{S^3} \wedge dx^{11},
   \nonumber \\
   r^2 &=& \sum_i(x^i)^2,   ~~~~ x^{11}= R_{11}\phi,
\label{metric} \eea where $\mu, \nu = 0, 1, \cdots, 5, ~~i, j =6, \cdots, 9$,  and $ 0 \leq \phi \leq 2\pi~ $, $dv_{S^4}$ denotes the volume form of a
unit $S^4$ and $l_p$ is the Planck length in the 11-dimensional theory. We would 
like to study the dynamics of a M2-brane in the above background and study the
homogeneous solutions only. The action for a single M2-brane is given by
\bea S_{M2} = -T_2 \int{d^3 \xi \sqrt{ -{\rm det} P[G]_{\mu\nu}} + T_2 \int P[A]},\eea 
where $T_2 = \frac{1}{4{\pi}^2{l_p}^3}$ is the tension 
of the M2-brane, $P[G]_{\mu \nu}$ and $P[A]$ are the pull back of the metric and the three form
field onto the worldsheet given by
\bea P[G]_{\mu\nu} &=& \frac{\partial X^{M}}{\partial \xi^{\mu}}\frac{\partial X^{N}}{\partial \xi^{\nu}} G_{MN} (X) \nonumber \\ 
P[A] &=& \frac{1}{6} \epsilon^{\mu\nu\rho} \frac{\partial X^{M}}{\partial \xi^{\mu}}\frac{\partial X^{N}}{\partial \xi^{\nu}}\frac{\partial X^{p}}{\partial \xi^{\rho}} A_{MNP} (X).
\eea

The indices $M, N, P$ runs over the 11-dimensional spacetime. We shall 
restrict ourselves to the case when the transverse directions of the M5-brane
depends only on time. So the dynamics will be governed only by the Nambu-Goto
part of the action. The action of the M2-brane is given by
\bea 
S_{M2} = - V T_2 \int{dt \sqrt{H^{-1} - \dot{X^i} \dot{X^i} - \dot{X^{11}} \dot{X^{11}}}} \ ,
\nonumber \\ \label{43}
\eea
where $V$ is the volume of the M2-brane.
The action (\ref{43}) of M2-brane in the background of 
stacks of M5-branes is given in terms of polar 
coordinates $X^6 = R \cos\theta, X^7 = R \sin\theta$ as
\begin{equation}
S_{M2} = -T_2V\int dt\sqrt{H^{-1}-{\dot{R}}^2-
R^2{\dot{\theta}}^2-{R_{11}}^2{\dot{\phi}}^2}
\nonumber \\ \label{44}
\end{equation}
The components of stress-energy tensor calculated from 
 (\ref{44}) are  
\begin{equation}
T_{00} =E = \frac{T_2}{H}\frac{1}
{\sqrt{H^{-1}-{\dot{R}}^2-
R^2{\dot{\theta}}^2-{R_{11}}^2{\dot{\phi}}^2}},
\nonumber \\ \label{45}
\end{equation}
\begin{equation}
T_{ij} =-T_2R^2
\sqrt{H^{-1}-{\dot{R}}^2-
R^2{\dot{\theta}}^2-{R_{11}}^2{\dot{\phi}}^2},
\nonumber \\ \label{46}
\end{equation}
\begin{equation}
T_{\phi\phi} = -T_2{R_{11}}^2
\sqrt{H^{-1}-{\dot{R}}^2-
R^2{\dot{\theta}}^2-{R_{11}}^2{\dot{\phi}}^2}.
\nonumber \\ \label{47}
\end{equation}
The angular momenta are given by
\begin{equation}
L_{\theta} = \frac{T_2R^2\dot{\theta}}
{\sqrt{H^{-1}-{\dot{R}}^2-
R^2{\dot{\theta}}^2-{R_{11}}^2{\dot{\phi}}^2}},
\nonumber \\ \label{48}
\end{equation}
and
\begin{equation}
L_{\phi} = \frac{T_2{R_{11}}^2\dot{\phi}}
{\sqrt{H^{-1}-{\dot{R}}^2-
R^2{\dot{\theta}}^2-{R_{11}}^2{\dot{\phi}}^2}}.
\nonumber \\ \label{49}
\end{equation}
From (\ref{45}) and (\ref{48}) we can solve for 
$\dot{\theta}$ as 
\begin{equation}
\dot{\theta} = \frac{L_{\theta}}{EHR^2}.
\nonumber \\ \label{50}
\end{equation}
From (\ref{45}) and (\ref{49}) and also from
(\ref{48}) and (\ref{49}), we get  $\dot{\phi}$ as
\begin{equation}
\dot{\phi} = \frac{L_{\phi}}{EH{R_{11}}^2} 
= \frac{L_{\phi}R^2}{L_{\theta}{R_{11}}^2}
\dot{\theta}.
\nonumber \\ \label{51}
\end{equation}
Using (\ref{50}) and (\ref{51}) in (\ref{45}), 
the equation for $\dot{R}$ reduces to
\begin{equation}
{\dot{R}}^2 = \frac{1}{H}-\frac{1}{E^2H^2}
\bigg({T_2}^2+\frac{{L_{\theta}}^2}{R^2} +
\frac{{L_{\phi}}^2}{{R_{11}}^2}\bigg).
\nonumber \\ \label{52}
\end{equation}
Defining ${T_e}^2={T_2}^2 +\frac{{L_{\phi}}^2}{{R_{11}}^2}$
as the effective tension, explained in \cite{Xu:2007wj}, 
we can rewrite (\ref{52}) as
\begin{equation}
{\dot{R}}^2 = \frac{1}{H}-\frac{1}{E^2H^2}
\bigg({T_e}^2+\frac{{L_{\theta}}^2}{R^2}\bigg).
\nonumber \\ \label{53}
\end{equation}
The above equation can be thought of as a particle 
of mass $(m=2)$ (with zero energy) moving in an effective 
potential of the form:  
\begin{equation}
V_{eff} = \frac{1}{E^2H^2}
\bigg({T_e}^2+\frac{{L_{\theta}}^2}{R^2}\bigg) -
\frac{1}{H}.
\nonumber \\ \label{54}
\end{equation}
Once again defining $\epsilon = \frac{E}{T_e}$ and 
$l = \frac{L_{\theta}}{T_e}$, the equation
(\ref{54}) becomes 
\begin{equation}
V_{eff} = \frac{1}{{\epsilon}^2H^2}
\bigg(1+\frac{l^2}{R^2}\bigg) -\frac{1}{H}.
\nonumber \\ \label{55}
\end{equation} 
In \cite{Xu:2007wj}, it was discussed that 
for $E > T_e$,  M2-brane escapes to 
infinity and for $E<T_e$, it falls into M5-branes.
From eqn (\ref{55}), it is clear that effective 
potential vanishes for particular values of 
energy density and angular momentum.
Thus for $l = \sqrt{\frac{N{l_p}^3}{R_{11}}}$ 
and $\epsilon = 1$, $V_{eff} = 0$. Hence, for this energy and angular momentum limits, the decay process of the 
M2-brane supressed and it becomes a stable object and 
revolves around the M5-branes.
The radial equation of motion derived from the action (\ref{44}) is given by:
\begin{equation}
\frac{d}{dt}\bigg(\frac{\dot{R}}{F(\dot{\theta},
\dot{\phi})}\bigg) = \frac{R}{F(\dot{\theta},
\dot{\phi})}\bigg({\dot{\theta}}^2 
- \frac{l^2}{H^2R^4}\bigg),
\nonumber \\ \label{56}
\end{equation}
where 
\begin{equation}
F(\dot{\theta},\dot{\phi}) =
\sqrt{H^{-1}-{\dot{R}}^2-
R^2{\dot{\theta}}^2-{R_{11}}^2{\dot{\phi}}^2}.
\nonumber \\ \label{57}
\end{equation}
In order to write the equation of motion only 
of radial component, we have to eliminate 
$\dot{\theta}$ and $\dot{\phi}$ in terms of $\dot{R}$ 
and $R$. Using (\ref{50}) and (\ref{51}) in (\ref{48}) and simplifing we get 
\begin{eqnarray}
{\dot{\theta}}^2 &=& \frac{{L_{\theta}}^2}
{{T_e}^2R^4}\Bigg(\frac{H^{-1} -{\dot{R}}^2}
{1+\frac{l^2}{R^2}}\Bigg) \cr & \cr
&=& \frac{l^2}{R^4H^2} - \frac{{\dot{R}}^2l^2}{R^4H}.
\nonumber \\ \label{58}
\end{eqnarray}
Again using (\ref{58}) in (\ref{51}) 
and  simplifying we get
\begin{equation}
{\dot{\phi}}^2 = \frac{{L_{\phi}}^2}
{{T_e}^2{R_{11}}^4}\Bigg(\frac{H^{-1} -{\dot{R}}^2}
{1+\frac{l^2}{R^2}}\Bigg).
\nonumber \\ \label{59}
\end{equation}
Now using (\ref{58}) and (\ref{59}) in (\ref{57}), 
we get
\begin{eqnarray}
F(\dot{\theta},\dot{\phi}) &=& \frac{T_1}{T_e}
\sqrt{\frac{H^{-1}-{\dot{R}}^2}{1+\frac{l^2}{R^2}}}
\cr & \cr
&=& \frac{F(0)}{\sqrt{1+\frac{l^2}{R^2}}},
\nonumber \\ \label{60}
\end{eqnarray}
where
\begin{equation}
F(0) = \frac{T_1}{T_e}\sqrt{H^{-1}-{\dot{R}}^2}.
\nonumber \\ \label{61}
\end{equation}
Finally using (\ref{58}) and (\ref{60}) in (\ref{56}), we can 
rewrite the equation of motion in terms of  radial coordinate as 
\begin{equation}
\frac{d}{dt}\Bigg(\frac{\dot{R}\sqrt{1+\frac{l^2}{R^2}}}
{F(0)}\Bigg) = -\frac{{\dot{R}}^2l^2}{R^3H}
\Bigg(\frac{\sqrt{1+\frac{l^2}{R^2}}}{F(0)}\Bigg).
\nonumber \\ \label{62}
\end{equation}
Similarly to the analysis given for F-string in 
$D5$-brane background, we can write a classically equvalent radial action 
of (\ref{44}) which gives the same equation of motion 
as (\ref{62}) and same energy density as (\ref{45}). 
The classically equivalent radial action is given by 
\begin{equation}
\tilde{S}_{M2} = -T_eV\int dt\sqrt{1+\frac{l^2}{R^2}}
\sqrt{H^{-1}-{\dot{R}}^2}.
\nonumber \\ \label{63}
\end{equation}
Note that this action is written in terms of effective 
tension and orbital angular momentum. The effect of 
other non-zero angular momentum is absorbed in the definition of $T_e$.

However the exact analogy with \cite{Jun:2009bn} 
is unknown because here 
we don't have a tachyon potential. So it remains to be seen whether this action gives more information about the radial dynamics of the membrane.
\section*{Acknowledgements} We thank K L Panigrahi for useful discussions. The 
work of RRN is supported by SERC, DST fast track project SR/FTP/PS-19/2009.

\end{document}